\documentclass[12pt]{article}

\usepackage[utf8]{inputenc}
\usepackage{geometry}
\usepackage{graphicx}
\usepackage[outdir=./]{epstopdf}
\usepackage{array}
\usepackage{subfig}
\usepackage{slashed}
\usepackage{amsmath}
\usepackage{amsfonts}
\usepackage{stmaryrd}
\SetSymbolFont{stmry}{bold}{U}{stmry}{m}{n} 
\usepackage{amssymb}
\usepackage[cm]{fullpage}
\usepackage{cite}
\usepackage{epsfig}
\usepackage{comment}
\usepackage{hyperref}
\usepackage{multirow}
\usepackage{cancel}
\usepackage{mathrsfs}

\usepackage{cleveref}
\crefname{section}{Â§}{Â§Â§}
\Crefname{section}{Â§}{Â§Â§}

\numberwithin{equation}{section}

\def\p{\partial}

\def\0{{(0)}}
\def\1{{(1)}}
\def\2{{(2)}}

\def\<{\langle }
\def\>{\rangle }

\newcommand{\bea}{\begin{eqnarray}}

\newcommand{\eea}{\end{eqnarray}}

\newcommand{\be}{\begin{equation}}
\newcommand{\ee}{\end{equation}}
\newcommand{\ba}{\begin{align}}
\newcommand{\ea}{\end{align}}

\makeatletter
  \let\over=\@@over \let\overwithdelims=\@@overwithdelims
  \let\atop=\@@atop \let\atopwithdelims=\@@atopwithdelims
  \let\above=\@@above \let\abovewithdelims=\@@abovewithdelims
\renewcommand\section{\@startsection {section}{1}{\z@}%
                                   {-3.5ex \@plus -1ex \@minus -.2ex}
                                   {2.3ex \@plus.2ex}%
                                   {\normalfont\large\bfseries}}

\renewcommand\subsection{\@startsection{subsection}{2}{\z@}%
                                     {-3.25ex\@plus -1ex \@minus -.2ex}%
                                     {1.5ex \@plus .2ex}%
                                     {\normalfont\bfseries}}
\makeatother

\DeclareMathOperator{\U}{U}
\DeclareMathOperator{\SU}{SU}

\newcommand{\beq}{\begin{equation}}
\newcommand{\eeq}{\end{equation}}
\newcommand{\beqa}{\begin{eqnarray}}
\newcommand{\eeqa}{\end{eqnarray}}
\newcommand{\beqar}{\begin{eqnarray*}}

\newcommand{\mc}[1]{\mathcal{#1}}

\def\[{\[}
\def\]{\]}

\def\mcc{{\mathbb C}}
\def\mrr{{\mathbb R}}
\def\mzz{{\mathbb Z}}

\newcommand{\bd}[1]{\begin{fmffile}{#1}\begin{fmfgraph*}}
\newcommand{\ed}{\end{fmfgraph*}\end{fmffile}}

\newcommand{\mN}{\mathcal{N}}
\newcommand{\N}{\mathcal{N}}
\newcommand{\mb}{\mathbb}

\newcommand{\Nstar}{\mN=2^\star}
\newcommand{\Pa}{\partial}
\newcommand{\Bp}{\bar\partial}

\begin{document}

\pagenumbering{Alph} 
\begin{titlepage}

\unitlength = 1mm~
\vskip 2cm
\begin{center}

{\Large{\textsc{Instanton Corrections for $m$ and $\Omega$}}}

\vspace{0.8cm}
Micha Moskovic\,{}\footnote{\tt moskovic@to.infn.it} and Ahmad Zein Assi\,{}\footnote{\tt zeinassi@cern.ch}

\vspace{1cm}

{\it  ${}^1$ Università di Torino, Dipartimento di Fisica and I.N.F.N. - sezione di Torino, Via P. Giuria 1, I-10125 Torino, Italy\\
	  ${}^2$ High Energy Section - ICTP, Strada Costiera, 11-34014 Trieste, Italy 

}

\vspace{0.8cm}

\begin{abstract}
In this paper, we study instanton corrections in the $\Nstar$ gauge theory by using its description in string theory as a freely-acting orbifold. The latter is used to compute, using the worldsheet, the deformation of the Yang-Mills action. In addition, we calculate the deformed instanton partition function, thus extending the results to the non-perturbative sector of the gauge theory. As we point out, the structure of the deformation is extremely similar to the $\Omega$-deformation, therefore confirming the universality of the construction. Finally, we comment on the realisation of the mass deformation using physical vertex operators by exploiting the equivalence between Scherk-Schwarz deformations and freely-acting orbifolds. 

\end{abstract}

\vspace{1.0cm}

\end{center}

\end{titlepage}

\pagenumbering{arabic} 

\pagestyle{empty}
\pagestyle{plain}

\def\vx{{\vec x}}
\def\p{\partial}
\def\po{$\cal P_O$}

\pagenumbering{arabic}

\tableofcontents

\section{Introduction}

One of the most difficult challenges of the past decades has been to obtain a full understanding of strongly coupled gauge theories. The lack of a general strategy has lead to considering toy models that can be `exactly' solved and, hence, increase our understanding of some aspects of more realistic theories. In this context, supersymmetric gauge theories play an essential role as shown by the work of Seiberg and Witten \cite{Seiberg:1994rs}. The low energy effective action is encoded in a single holomorphic function, the prepotential, which can be computed directly through equivariant localisation \cite{Losev:1997wp,Nekrasov:2002qd,Nekrasov:2003rj}. The latter is based on the so-called $\Omega$-deformation which regularises the integrals over the instanton moduli space.

Independently, ever since its construction, topological string theory has proven fruitful in bridging different branches of mathematics and physics. In particular, despite its topological nature, it encodes physical observables in string theory as higher derivative couplings in the effective action \cite{Bershadsky:1993cx,Antoniadis:1993ze}. The fascinating, yet mysterious connection to supersymmetric gauge theories is the fact that the latter observables are nothing but higher-derivative gravitational corrections to the prepotential. More precisely, the $\Omega$-deformation is, from the string point of view, given by the vacuum expectation values of the graviphoton \cite{Bershadsky:1993cx,Antoniadis:1993ze} and a particular gauge field strengths \cite{AFHNZ,Antoniadis:2013mna}. Therefore, the string theory picture is crucial to interpret gauge theory deformations more physically, and this gauge/string interplay has induced numerous discoveries on both sides.

In this context, the mass deformation of $\N=4$ gauge theories plays a very special role. Not only it is a deformation of gauge theory that carries a clear physical meaning, it also triggers an RG flow between $\N=4$ and $\N=2$ gauge theories which still preserves exact marginality of the gauge coupling and, hence, the (quasi-)modular properties of the prepotential and the physical observables \cite{Billo:2014bja,Billo':2015ria,Billo':2015jta,Billo:2016zbf,Ashok:2016ewb}.  From the string theory perspective, it can be implemented as a freely-acting orbifold of $\N=4$ compactifications of \emph{any} string theory \cite{Florakis:2015ied} in a universal fashion. This has been also checked in \cite{Florakis:2015ied} by coupling the worldsheet CFT to the self-dual $\Omega$-deformation using graviphotons and reproducing the $\Nstar$ gauge theory partition function \cite{Nekrasov:2003rj,Billo:2013fi,Billo:2014bja}.

Following this picture, the natural question of the existence of similar supersymmetric deformations can be posed. In this work, we argue that there are none and we attempt at unifying all deformations from the orbifold picture, while pointing out the main differences between them. From the gauge theory point of view, the $\Omega$-deformation can be geometrically implemented by a dimensional reduction from the $\N=1$ theory in six dimensions to four on a $T^2$ with twisted boundary conditions \cite{Nekrasov:2003rj}, and this can be straightforwardly lifted to string theory \cite{Hellerman:2011mv,Orlando:2013yea}. The geometric picture is, hence, convenient to unify different gauge theory deformations. This is precisely the path we follow by exploiting the exactness of the underlying string background.

In the present work, we first focus on the mass deformation of the pure $\N=4$ gauge theory by implementing it geometrically in a D-brane setup presented in Section \ref{FluxMass}. Furthermore, we derive in Section \ref{YangMills} the mass-deformed super-Yang-Mills action from string theory by using the exact description of the freely-acting orbifold \cite{Florakis:2015ied}. Using D-instantons, this is extended to the instanton sector in Section \ref{ADHM} in which we comment on the similarities with the $\Omega$-deformation. Section \ref{Conclusion} contains our conclusions, and some useful technical results are summarised in Appendix \ref{appendix:spinors}. Finally, in Appendix \ref{appendix:vertex}, we use the flux equivalent of the orbifold \cite{Ferrara:1987qp,Ferrara:1988jx,Antoniadis:1998ep,Condeescu:2012sp} in order to describe the mass deformation as a vertex operator. 


\section{The string description of the mass deformation}\label{FluxMass}

\subsection{A freely acting orbifold}

We first review the string theory background that gives rise to the $\Nstar$ mass deformation in the field theory limit \cite{Florakis:2015ied}.
In order to realise the $\Nstar$ gauge theory from a string theory construction, one first needs to start from a description of the $\mN=4$ theory and then break supersymmetry to $\mN=2$ by turning on a mass for the adjoint hypermultiplet.
Hence, consider the standard setup consisting of a stack of $N$ parallel D3-branes in type IIB string theory whose low-energy worldvolume dynamics lead to the $\mN=4$ $\U(N)$ gauge theory.
In order to break supersymmetry in string theory, a simple way is to replace the space transverse to the D-branes by an orbifold.
Instead of $\mcc^3$, one might take $\mcc\times \mcc^2/\mzz_k$, where the identification is generated by
\begin{equation}
  (x,\hat z^1,\hat z^2) \sim (x,e^{\frac{2\pi i}{k}}\hat z^1, e^{-\frac{2\pi i}{k}}\hat z^2) \, .
  \label{eq:stdorbifold}
\end{equation}
As the $\mzz_k$ orbifold group acts on space-time by a subgroup of $\SU(2)$, the space-time has $\SU(2)$ holonomy and the background preserves half of the supercharges, corresponding to an $\mN=2$ gauge theory on the D3-brane worldvolume.
While it has the correct amount of supersymmetry, this gauge theory is not $\Nstar$, but instead the $\hat A_{k-1}$ quiver gauge theory, with gauge group $\U(N)^k$ and bifundamental hypermultiplets\cite{Douglas:1996sw}. The stringy origin of this product gauge group is related to the fact that the orbifold action is not free: the origin of the $(\hat z^1, \hat z^2)$ space is fixed, and a regular D-brane placed at this orbifold singularity can split into its fractional brane constituents, each type of fractional brane being associated to a factor of the gauge group.

This suggests that, by taking instead a freely acting orbifold, which does not have any orbifold singularity, we keep a gauge group with a single factor and may hence engineer the $\Nstar$ theory.
We thus consider type IIB string theory on the background $\mrr^{3,1}\times\mc{M}$ where the space $\mc{M}$ is the orbifold of $\mcc^3$ obtained by identifying
\begin{align}
  (x,\hat z^1,\hat z^2) &\sim (x+2\pi R_1, e^{\frac{2\pi i}{k}}\hat z^1, e^{-\frac{2\pi i}{k}}\hat z^2) \label{eq:orbifold1} \\
  &\sim (x+2\pi i R_2, e^{\frac{2\pi i}{k}}\hat z^1, e^{-\frac{2\pi i}{k}}\hat z^2) \, .
  \label{eq:orbifold2}
\end{align}
Compared to the non-freely acting orbifold \eqref{eq:stdorbifold}, the rotation in the $(\hat z^1,\hat z^2)$ space is now accompanied by translations along the real or imaginary axes of the $x$ space which hence describes a two-torus $T^2$ in complex coordinates. When restricting to the $(\hat z^1,\hat z^2)$ space, the origin is still a fixed point, but this point always gets translated along the torus so there is no fixed point in the full internal space $\mc{M}$.

One can now place D-branes in this background in order to engineer the gauge theory degrees of freedom \cite{Takayanagi:2001aj}. We consider a stack of $N$ D5-branes, wrapping $\mrr^{3,1}$ and the $T^2$ inside $\mc{M}$, and localised at $\hat z^1=\hat z^2=0$, which are allowed boundary conditions since they are invariant under \eqref{eq:orbifold1}, \eqref{eq:orbifold2}.
In a hereafter prescribed $\alpha'\to 0$ limit, the worldvolume dynamics on the D-branes reduces to a four-dimensional gauge theory. Note that we could equivalently consider a system of D3-branes by applying the appropriate T-dualities.

Quantisation of the open strings with the boundary conditions we chose proceeds as usual. One only needs to pay attention to impose that the vertex operators be gauge-invariant, that is invariant under the identifications \eqref{eq:orbifold1}, \eqref{eq:orbifold2}.
Consider the charge $Q$ corresponding to the $\U(1)\supset \mzz_k$ rotations that appear in the orbifold construction.
For the ``fundamental" zero-momentum vertex operators, one has
\begin{align}
  &[Q,\partial \hat Z^1]=[Q,\hat\psi^1]=[Q,\bar \partial \hat Z^1]=[Q,\hat{\tilde\psi}^1]=[Q, \partial \hat{\bar Z}^2]=[Q,\hat{\bar\psi}^2]=[Q, \bar \partial \hat{\bar Z}^2]=[Q,\hat{\tilde{\bar\psi}}^2]=1 \, , \nonumber \\
  &[Q,\partial \hat Z^2]=[Q,\hat\psi^2]=[Q,\bar \partial \hat Z^2]=[Q,\hat{\tilde\psi}^2]=[Q, \partial \hat{\bar Z}^1]=[Q,\hat{\bar\psi}^1]=[Q, \bar \partial \hat{\bar Z}^1]=[Q,\hat{\tilde{\bar\psi}}^1]=-1 \, , \nonumber \\
  &[Q, \partial X]=[Q,\chi]=[Q, \bar \partial X]=[Q,\tilde \chi]=0 \, ,
  \label{eq:Qvertex}
\end{align}
where we have also written to the right of every bosonic operator the worldsheet fermion in the same supermultiplet.
For operators without momentum along $T^2$, this charge detects invariance under the orbifold identifications: such operators are invariant if and only if they have zero charge.
More generally, a zero-momentum vertex operator $O_j$ with charge $j$ under $Q$, dressed with momentum along $T^2$,
\begin{equation}
  O_j e^{\frac{i}{2}(p X + \bar p \bar X)} = e^{\frac{2\pi i j}{k}} O_j e^{\frac{i}{2}(p (X+2\pi R_1) + \bar p (\bar X+2\pi R_1))} = e^{\frac{2\pi i j}{k}} O_j e^{\frac{i}{2}(p (X+2\pi iR_2) + \bar p (\bar X-2\pi iR_2))} \, ,
\end{equation}
is gauge-invariant if
\begin{equation}
  \frac{j}{k}+\frac{1}{2}(p+\bar p) R_1 \in \mzz \, , \quad \frac{j}{k} + \frac{i}{2} (p-\bar p) R_2 \in \mzz \, .
\end{equation}
These conditions are equivalent to the following quantisation conditions on the complex momentum $p$ along $T^2$:
\begin{equation}
  p = \left(\frac{n_1}{R_1}-\frac{j}{kR_1}\right)+i\left(\frac{n_2}{R_2}-\frac{j}{kR_2}\right) \, , \quad n_1, n_2 \in \mzz \, .
  \label{eq:pquanti}
\end{equation}
The first terms in the brackets are the standard quantisation of the momentum due to the compactness of the torus, which now receive additional shifts due to the presence of the orbifold.
These shifts are the only effect of the orbifold that is visible in the open-string sector we are concerned with.
Unlike a non-freely acting orbifold which projects out the non-invariant operators from the spectrum, in the freely acting setting any non-invariance of the zero-momentum part of a vertex operator can be compensated by a shift of its quantised momentum.

The quantisation condition \eqref{eq:pquanti} gives a four-dimensional mass formula that explicitly depends on the $Q$-charge of the state,
\begin{equation}
  M^2_{\text{4D}}=\left(\frac{n_1}{R_1}-\frac{j}{kR_1}\right)^2+\left(\frac{n_2}{R_2}-\frac{j}{kR_2}\right)^2 + \cdots \, ,
  \label{eq:m4D}
\end{equation}
where the $\cdots$ stand for additional contributions for higher mass levels of the string that are not affected by the presence of the orbifold, and are irrelevant in the field theory limit.
In the absence of the orbifold, the massless sector would have $n_1=n_2=0$ and be made of the full $\mc{N}=4$ vector multiplet.
All states in this multiplet have $|j|\le1$ and, while those with $j=0$ remain massless, those with $|j|=1$ become massive, with
\begin{equation}
  M^2=|m_{\textrm{4D}}|^2 \, , \quad \text{where} \quad m_{\textrm{4D}}=\frac{1}{kR_1}-\frac{i}{kR_2} \, .
  \label{eq:m2}
\end{equation}
The massive states can easily be seen to fit into an $\mc{N}=2$ hypermultiplet, whereas the $\mc{N}=2$ vector multiplet states remain massless.
This mass spectrum is precisely the one of the $\mc{N}=2^\star$ theory.

Up to now, we have been quite imprecise with the exact limit that needs to be taken in order to engineer a four-dimensional gauge theory, as opposed to a higher dimensional theory or a string theory. In order to decouple the stringy effects and obtain a field theory, one needs to take $\alpha'\to0$. Simultaneously, one needs the mass of the hypermultiplet to stay finite but parametrically lighter than the excited string states, which requires
\begin{equation}
  0<\frac{1}{kR_a}\ll\frac{1}{\sqrt{\alpha'}} \, , \quad a = 1,2 \, .
  \label{eq:masslimit}
\end{equation}
These conditions can be solved by taking $k\gg 1$ and $\sqrt{\alpha'}/R_a\ll1$ while $\alpha',R_a\to0$\footnote{In principle, one can keep $k$ and $R_a$ arbitrary and the field theory limit is properly realised. However, if we want to obtain strictly a four-dimensional theory, one needs to send the radii of the torus to zero. Hence, we take the formal large $k$ limit such that the mass is finite and continuously interpolating between the $\N=2$ and $\N=4$ theories.}.

\subsection{The Melvin background}

The description of the mass deformation as a freely-acting orbifold, while appealing from the worldsheet string theory point of view as it is an exact CFT, is quite exotic from the target space point of view.
In particular, there should exist a smooth $m\to0$ limit that reduces the $\mc{N}=2^\star$ theory to $\mc{N}=4$, and one should be able to treat a small mass parameter as a perturbation of the flat background yielding the maximally supersymmetric theory.
We now present such a description based on the known equivalence of this class of orbifolds with supergravity backgrounds of the Melvin type \cite{Tseytlin:1994ei}.  

We perform the change of coordinates
\begin{align}
  z^1 &= e^{-\frac{i}{2}(mx+\bar m\bar x)}\hat z^1 \, , \nonumber \\
  z^2 &= e^{+\frac{i}{2}(mx+\bar m\bar x)}\hat z^2 \, \, ,
  \label{eq:zzhat}
\end{align}
where $m$ is given by \eqref{eq:m2}.
The new coordinates $z^1$ and $z^2$ have the property that they are invariant under the orbifold identifications \eqref{eq:orbifold1}, \eqref{eq:orbifold2}. One simply has
\begin{equation}
  (x,z^1,z^2)\sim (x+2\pi R_1, z^1, z^2) \sim (x+2\pi i R_2, z^1, z^2) \, ,
  \label{eq:unorbifold}
\end{equation}
and $\mc{M}$ is now seen to be topologically equivalent to $T^2\times \mcc^2$.
The price to pay is the appearance of a non-flat metric on $\mc{M}$,
\begin{align}
  ds^2_{\mc{M}}&= G_{IJ}(x^K)dx^Idx^J = dx d\bar x + d\hat z^1 d\hat{\bar z}^1 + d\hat z^2 d\hat{\bar z}^2 \nonumber \\
  &= dx d\bar x + dz^1d\bar z^1 + dz^2d\bar z^2 \nonumber \\
  &\quad - \frac{im}{2}(\bar z^1 dz^1 - z^1d\bar z^1 - \bar z^2 dz^2 + z^2d\bar z^2)dx - \frac{i\bar m}{2}(\bar z^1 dz^1 - z^1d\bar z^1 - \bar z^2 dz^2 + z^2d\bar z^2)d\bar x \nonumber \\
  &\quad + \frac{1}{4}(|z^1|^2+|z^2|^2)(mdx + \bar m d\bar x)^2 \, ,
  \label{eq:metric}
\end{align}
writing $x^I, I=1,\ldots,6$ for the coordinates on $\mc{M}$.
We have separated the contributions of different orders in the mass: the second line is the undeformed flat metric, the third line is a linear correction whereas the last line is a quadratic one.
By construction, this background (when supplemented with a Minkowski metric on the omitted $\mrr^{3,1}$ factor) is automatically a solution to type IIB supergravity preserving 16 supercharges.
Instead of the $R_1,R_2\to0$ limit we need to take, it is more natural in the supergravity description to have a decompactification limit $\tilde R_1,\tilde R_2\to\infty$.
This can be achieved by T-dualising on both cycles of the $T^2$, which turns on a $B$-field and a dilaton in addition to a non-trivial metric and yields a background of the fluxtrap type \cite{Hellerman:2011mv,Orlando:2013yea}.
This is not needed for our purposes, however, as we use this background to compute string theory amplitudes, which are insensitive to the chosen T-duality frame.

On the string worldsheet, we need to perform the same change of variables \eqref{eq:zzhat}. The result is that, instead of the freely acting orbifold CFT we started with, we now have a $\sigma$-model with target space $\mc{M}$ and metric given by \eqref{eq:metric}.
In order to compute amplitudes in the $\sigma$-model, we can work perturbatively in $m$ and compute amplitudes in the free theory with insertions of graviton vertex operators.
Splitting $G_{IJ}(x)=\delta_{IJ}+ \delta G_{IJ}(x)$ (the latter given by the two last lines of \eqref{eq:metric}), the $\sigma$-model action splits as $S=S_0+\delta S$, where $S_0$ is the free type II action and $\delta S$ is a $\sigma$-model with metric $\delta G_{IJ}$,
\begin{equation}
  \delta S = \frac{1}{2\pi}\int d^2zd^2\theta \delta G_{IJ}(X^K) DX^I\bar DX^J \, .
  \label{eq:deltaS}
\end{equation}
Here, $D=\partial_\theta + \theta \partial$, $\bar D=\partial_{\bar\theta} + \bar\theta \bar\partial$ are the worldsheet superderivatives and $X^I(z,\bar z,\theta,\bar\theta)=X^I + i\theta\psi^I + i\bar\theta\tilde\psi^I + \theta\bar\theta F^I$ is the worldsheet superfield associated to the coordinate $x^I$. We refer the reader to Appendix \ref{appendix:vertex} for more details on this approach. In what follows, we derive the gauge theory effective actions based on the exact CFT description.

\section{Deformed Yang-Mills from the worldsheet}\label{YangMills}

\subsection{\texorpdfstring{Review of the $\N=4$ setup}{Review of the N=4 setup}}

As noted previously, we realise the $SU(N)$ gauge theory through a system of $N$ D5-branes in a type IIB orientifold. In the simplest case of toroidal compactification on $T^6$, this leads to the maximally supersymmetric case of $\mc N=4$ $SU(N)$ gauge theory. Here, we consider a decompactified version of the torus as $\mb C^3$ such that $N$ can be generic. The Euclidean Lorentz group is broken,
\begin{equation}\label{LorentzDecomp}
 SO(10)\rightarrow SO(4)\times SO(6)\,
\end{equation}
whose covering group is $\mathfrak{G}=SU(2)_{L}\times SU(2)_R\times SU(4)$. Note that $SU(4)$ plays the role of the R-symmetry for the $\mc N=4$ theory. The ten-dimensional index $M\in\llbracket0,9\rrbracket$ is decomposed into longitudinal and transverse directions with respect to the space-time:
\begin{equation}
 M\rightarrow(\mu,I)\in\llbracket0,3\rrbracket\times\llbracket4,9\rrbracket\,.
\end{equation}
The string coordinates obey specific boundary conditions depending on the location of its endpoints and, in particular, these are Neumann along the longitudinal direction of the D5-branes. The ten-dimensional spin field preserved by the GSO projection is decomposed by \eqref{LorentzDecomp} as
\begin{equation}
 S\rightarrow(S_{\alpha}S_{A},S^{\dot\alpha}S^{A})\,,
\end{equation}
where $\alpha,\dot\alpha$ denote chiral, anti-chiral spinors in four dimensions and the upper, lower $A$ index refers to fundamental, anti-fundamental representations of $SO(6)$, see Appendix \ref{appendix:spinors} for more details. 

Let us now discuss the massless spectrum obtained from the open string sector and which comes from the reduction of the ten-dimensional $\mc N=1$ multiplets. The endpoints of the open string can be located on the D5-branes and this is dubbed the 5-5 or gauge sector. The massless excitations consist of a number of $\mathcal{N}=4$ vector multiplets, each of which containing a vector field $A^\mu$, six real scalars $\phi^I$, as well as two gaugini $(\Lambda^{\alpha A}, \Lambda_{\dot\alpha A})$ which transform in the $(\mathbf{2},\mathbf{1},\mathbf{4})\oplus(\mathbf{1},\mathbf{2},\mathbf{\bar 4})$ representation of $\mathfrak{G}$. The bosonic degrees of freedom stem from the NS sector, while the fermionic ones from the R sector. This sector taken separately realises an $\mc N=4$ super-Yang-Mills theory living on the four-dimensional space-time.

The vertex operators for these massless fields are
\begin{align}
 V_{A}(z)&= \frac{A_{\mu}(p)}{\sqrt{2}}\psi^{\mu}(z)\,e^{ip\cdotp X(z)}\,e^{-\varphi(z)}\,,&V_{\phi}(z)&= \frac{\phi_I(p)}{\sqrt{2}}\psi^I(z)\,e^{ip\cdotp X(z)}\,e^{-\varphi(z)}\,,\label{AVop}\\
 V_{\Lambda}(z)&= \Lambda^{\alpha A}S_{\alpha}(z)S_{A}(z)\,e^{ip\cdotp X(z)}\,e^{-\tfrac{1}{2}\varphi(z)}\,,&V_{\bar\Lambda}(z)&= \Lambda_{\dot\alpha A}S^{\dot\alpha}(z)S^{A}(z)\,e^{ip\cdotp X(z)}\,e^{-\tfrac{1}{2}\varphi(z)}\label{LambdaVop}\,.
\end{align}
Here, we have set $2\pi\alpha'=1$. In order to recover the usual dimensions, one must rescale the fields with $(2\pi\alpha')^{\frac{3-2\nu}{4}}$ and, as usual, $\nu_{\textrm{R,NS}}=0,\frac{1}{2}$.


\subsection{Mass deformation}

We first focus on the bosonic mass terms. Due to the very nature of the orbifold action, we split the scalar vertex operator of the $\mN=4$ vector multiplet according to $\mb C^3\rightarrow \mb C\times\mb C^2$ as\footnote{Recall that one may think of $\mb C$ as a square torus with radii $R_1$ and $R_2$.}
\begin{align}
 V_{\chi}(z,p)&=\frac{\chi(p)}{\sqrt{2}}\psi^3(z)e^{ip\cdot Z(z)}e^{-\varphi(z)}\sim\frac{\chi(p)}{\sqrt{2}}\left(\Pa X-i(p\cdot\psi)\psi^3\right)e^{ip\cdot Z(z)}\,,\\
 V_{k}(z,p)&=\frac{\phi_k(p)}{\sqrt{2}}\psi^k(z)e^{ip\cdot Z(z)}e^{-\varphi(z)}\sim\frac{\phi_k(p)}{\sqrt{2}}\left(\Pa Z^k-i(p\cdot\psi)\psi^k\right)e^{ip\cdot Z(z)}\,,
\end{align}
where $\varphi$ counts the superghost number. More precisely, it is easy to see that under the decomposition $\mN=4\rightarrow\mN=2$, $\chi$ is the scalar of the $\mN=2$ vector multiplet while $\phi_k$ belongs to a hypermultiplet.
Following the analysis of \cite{Florakis:2015ied} reviewed in Section \ref{FluxMass}, in the standard $\mc N=2$ theory, the hypermultiplet states $\phi_k$ are odd under the orbifold projection and, therefore, are projected out of the spectrum as can be easily seen from their vertex operators. However, in the freely-acting orbifold case, they become invariant under the orbifold projection at the cost of acquiring a mass given by \eqref{eq:m2}. From the vertex operator, this can be seen from the rational shifts in the momenta of the states. In particular, the scalar $\chi$ remains massless.
The fermionic states follow by supersymmetry. More precisely, splitting the $\N=4$ fermions according to the $SO(6)\rightarrow SO(2)\times SO(4)$ decomposition, one easily sees that the hyperini are projected in the spectrum while acquiring a Scherk-Schwarz mass.

In order to complete the perturbative analysis, we check that the correct trilinear coupling of the scalar fields stemming from the superpotential is reproduced. To see this, one way is to calculate the coupling of three scalar fields to first derivative order (in the internal momentum along the shift direction). That is
\begin{align}
 \left\<\left\<V_{\chi}(x_1)V_k(x_2)V_k(x_2)\right\>\right\>
\end{align}
and extract the linear term in $p_3$. This calculation is staightforward and leads to
\begin{equation}
 ip_3\left[\phi_i\phi_j\delta^{ij}\chi-\phi_i\chi\phi_l\delta^{il}\right]\,.
\end{equation}
One must keep in mind that the fields are in the adjoint representation of the gauge group. Therefore, we obtain a non-trivial trilinear coupling of the form
\begin{equation}
 m\,\phi_i\left[\phi_j,\chi\right]\,,
\end{equation}
and similarly for $\bar m$. All other possible couplings are either zero or suppressed in the field theory limit. Consequently, we obtain the correct mass-deformed Yang-Mills action
\begin{align}
 \mc L_{\textrm{SYM}}&=\frac{1}{g_{\textrm{YM}}^2}\textrm{Tr}\left\{\frac{1}{2}F^2-2\bar\Lambda_{\dot\alpha A}\bar{\slashed{D}}^{\dot\alpha\beta}{\Lambda_{\beta}}^A+D_{\mu}\phi^a\,D^{\mu}\bar\phi_a-\frac{1}{2}[\phi_a,\phi_b]^2+2i(\bar\Sigma^a)_{AB}\Lambda^{\alpha A}[\phi_a,{\Lambda_{\alpha}}^{B}]\right.\nonumber\\
&+\left.\tfrac{1}{2}|m|^2|\phi_k|^2-m\,\bar\chi\left(\left[\phi_1,\bar\phi_1\right]-\left[\phi_2,\bar\phi_2\right]\right)+2\,i\,m\,P_{AB}\Lambda^{\alpha A}{\Lambda_{\alpha}}^B+\textrm{c.c}\right\}\,.\label{SYMaction}
\end{align}
Notice that we have written the fermionic mass term covariantly in terms of the projector $P$ which has the precise $\Sigma$-matrix structure $\Sigma^{1\bar13}-\Sigma^{2\bar23}$ that keeps the fermions of the $\N=2$ vector multiplet massless.


\section{Probing the instanton sector}\label{ADHM}

\subsection{ADHM instantons}

In the D5-brane realisation of the gauge theory, we add an arbitrary number of D1-branes wrapping the $T^2$ in order to describe the gauge theory instantons. Indeed, the massless excitations of the open strings with at least one endpoint on a D-instanton correspond to the ADHM moduli. Hence, there are two classes of open string excitations in the instanton sector, depending on whether only one endpoint of the open string lies on a D-instanton (5-1, 1-5 or mixed sector) or both (1-1 or unmixed sector). The ADHM moduli are non dynamical fields due to the instantonic nature of the D1-branes. Indeed, the states in this sector cannot carry any momentum because of the Dirichlet boundary conditions in all transverse directions\footnote{Recall that a four-dimensional limit is taken by shrinking the torus to zero size.}.

In the 1-1 and NS sector, we have ten bosonic moduli that can be written as a real vector $a^\mu$ and six scalars $\chi^I$. From the point of view of the gauge theory living on the world-volume of the D5-branes, $a^\mu$ corresponds to the position of gauge theory instantons. In the Ramond sector, there are sixteen fermionic moduli denoted by $M^{\alpha A}$, $\lambda_{\dot\alpha A}$. This unmixed sector is the close parallel to the gauge sector as can also be seen from their vertex operators:
\begin{align}
 V_{a}(z)&= g_0\,a_{\mu}\psi^{\mu}(z)e^{-\varphi(z)}\,,&V_{\chi}(z)&= \frac{\chi_I}{\sqrt{2}}\psi^I(z)e^{-\varphi(z)}\,,\label{aVop}\\
 V_{M}(z)&= \frac{g_0}{\sqrt{2}}\,M^{\alpha A}S_{\alpha}(z)S_{A}(z)e^{-\tfrac{1}{2}\varphi(z)}\,,&V_{\lambda}(z)&= \lambda_{\dot\alpha A}\,S^{\dot\alpha}(z)S^{A}(z)\,e^{-\tfrac{1}{2}\varphi(z)}\,.\label{MlambdaVop}
\end{align}
Here, $g_0$ is the D(-1)-brane coupling constant. For a Dp-brane, it is defined as
\begin{equation}
 g_{p+1}^2=4\pi(2\pi\sqrt{\alpha'})^{p-3}\,g_s\,.
\end{equation}

As for the mixed moduli, from the NS sector, the fermionic coordinates give rise to two Weyl spinors of $SO(4)$ and are called $(\omega_{\dot\alpha},\bar\omega_{\dot\alpha})$. The latter have the same chirality owing to the specific choice of boundary conditions of the D1-branes. From the gauge theory point of view, these fields describe the size of the instanton. In the R sector, we have two Weyl fermions $(\mu^A, \bar\mu^A)$ transforming in the fundamental representation of $SO(6)$.

\begin{table}[ht]
\begin{center}
\begin{tabular}{|c||c||c|c|}\hline
&&&\\[-11pt]
\bf{Sector} & \bf{Field} & \bf{Statistic} & \bf{R / NS} \\\hline\hline
&&&\\[-11pt]
\multirow{7}{*}{5-5} & $A^\mu$ & boson & NS \\
&&&\\[-11pt]\cline{2-4}
&&&\\[-11pt]
 & $\Lambda^{\alpha A}$ & fermion & R \\
 &&&\\[-11pt]\cline{2-4}
 &&&\\[-11pt]
 & $\Lambda_{\dot{\alpha} A}$ & fermion & R \\
 &&&\\[-11pt]\cline{2-4}
 &&&\\[-11pt]
 & $\phi^I=\chi,\phi^k$ & boson & NS \\ &&&\\[-11pt]\hline\hline
 &&&\\[-11pt]
 \multirow{7}{*}{1-1} & $a^\mu$ & boson & NS \\
&&&\\[-11pt]\cline{2-4}
&&&\\[-11pt]
 & $\chi^I=\phi,\chi^k$ & boson & NS \\
&&&\\[-11pt]\cline{2-4}
&&&\\[-11pt]
& $M^{\alpha A}$ & fermion & R \\
&&&\\[-11pt]\cline{2-4}
&&&\\[-11pt]
& $\lambda_{\dot{\alpha} A}$ & fermion & R \\
&&&\\[-11pt]\hline\hline
&&&\\[-11pt]
\multirow{3}{*}{1-5/5-1} & $\omega_{\dot{\alpha}},\bar\omega_{\dot{\alpha}}$ & boson & NS \\
&&&\\[-11pt]\cline{2-4}
&&&\\[-11pt]
 & $\mu^A$ & fermion & R \\[2pt]\hline
\end{tabular}
\end{center}
\caption{Summary of the massless spectrum of the D1/D5 system and the decomposition of the scalar fields.}
\label{Tab:Fields}
\end{table}

The vertex operators for the mixed sectors contain the so-called \emph{twist operators} that implement a change in the coordinates boundary conditions from Dirichlet to Neumann and vice-versa. These are bosonic fields denoted $\Delta,\bar\Delta$ and carry conformal dimension 1/4. The vertex operators are
\begin{align}
 V_{\omega}(z)=& \frac{g_0}{\sqrt{2}}\,\omega_{\dot\alpha}\Delta(z)S^{\dot\alpha}(z)e^{-\varphi(z)}\,,& V_{\bar\omega}(z)=& \frac{g_0}{\sqrt{2}}\,\bar\omega_{\dot\alpha}\bar\Delta(z)S^{\dot\alpha}(z)e^{-\varphi(z)}\,\label{OmegaVop}\\
 V_{\mu}(z)=& \frac{g_0}{\sqrt{2}}\,\mu^A\Delta(z)S_{A}(z)e^{-\frac{1}{2}\varphi(z)}\,,&V_{\bar\mu}(z)=& \frac{g_0}{\sqrt{2}}\,\bar\mu^{A}\bar\Delta(z)S_{A}(z)e^{-\frac{1}{2}\varphi(z)}\,.\label{muVop}
\end{align}

\subsection{\texorpdfstring{Mass terms and $\Omega$}{Mass terms and Omega}}

We now focus on the analysis of the mass terms which is very similar to the perturbative one. Indeed, from the quantisation of the exact CFT, one immediately finds that some of the massless excitations of the open strings with endpoints on the D-instantons acquire a Scherk-Schwarz mass due to the non-trivial boundary conditions. As mentioned previously, in the absence of the shift, these states are simply projected out of the spectrum.

More precisely, in the $1$-$1$ sector, decomposing the scalars $\chi^a$ as $\phi,\chi^k$, the fields $\chi^k$ acquire a mass while $\phi$ remains massless. Similarly, only the fermionic partners of $\chi^k$ do acquire a mass. These terms are the exact analogues of the perturbative ones. However, the instanton sector contains more states, namely the mixed 3-1 and 1-3 ones. It is easy to see that only the fermionic mixed states can acquire a mass since they carry an index under $SO(6)$. Indeed, under the freely-acting orbifold action, the $SO(4)$ part of them becomes massive, similarly to what happen for the unmixed fermions. Therefore, we obtain the correct mass-deformed part of the ADHM action:
\begin{align}
 \mc L_{\textrm{ADHM}}=&-\textrm{Tr}\left\{\left([\chi^\dag,\phi_{a\dot b}]+\bar m(\sigma^3\cdot\phi)_{a\dot b}\right)\left([\chi,\phi^{\dot ba}]+m(\sigma^3\cdot\phi)^{\dot ba}\right)\right\}\nonumber\\
&+2i\,\textrm{Tr}\left\{M^{\alpha A}\left(\left[\left(\bar\Sigma^a\right)_{AB}\chi_a,{M_{\alpha}}^B\right]+m\,P_{AB}{M_{\alpha}}^B\right)+\left(\left(\bar\Sigma^a\right)_{AB}\chi_a+m\,P_{AB}\right)\bar\mu^A\mu^B+\textrm{c.c}\right\}\,.\label{ADHMaction}
\end{align}
Here, we have used the $SO(4)$ spinor notation in order to write the action in an elegant fashion:
\begin{equation}
\phi_{a\dot b}\equiv \phi_k\left(\sigma^k\right)_{a\dot b}\,,
\end{equation}
see Appendix \ref{appendix:spinors} for more details. Notice that this effective action shares many features with the $\Omega$-deformed one \cite{Antoniadis:2013mna}. Indeed, the latter can be obtained by a similar freely-acting orbifold construction where the twist acts on the four-dimensional space-time where the gauge theory lives rather than the internal space. Using this picture, similarly to the fact that the D-brane instantons are localised at the orbifold point of the internal space in the mass-deformed theory, in the $\Omega$-deformed case they are localised at the origin of space-time. This leads to the fact that the $\Omega$-deformation regularises the path integral over the instanton moduli space and provides a way to directly calculate the Seiberg-Witten prepotential.
Moreover, since the anti-holomorphic terms in the $\Omega$-deformation are, in the ADHM action, $Q$-exact terms, the same is to be expected in the mass-deformed case.

Furthermore, similarly to the $\Omega$ case \cite{Billo:2006jm}, the action can be written as a $Q$-exact quantity, where $Q$ is the supercharge preserved by the D-brane system. The action of $Q$ would depend on $m$ holomorphically such that the mass-deformed instanton partition function depends holomorphically on the mass, as it is known. Therefore, the non-holomorphic terms in the effective action can be ignored as they decouple from physical quantities through $Q$-exactness. In this sense, similarly to what is done for $\Omega$, \emph{i.e.} realising it as a holomorphic constant background for physical string states, one can `truncate' the mass-deformed action to the holomorphic linear order, see Appendix \ref{appendix:vertex} for a more detailed discussion.


\section{Conclusions}\label{Conclusion}

In this work, we have studied in depth the string theory realisation of the $\Nstar$ gauge theory \cite{Florakis:2015ied} in terms of freely-acting orbifolds of $\N=4$ compactifications. Indeed, by adding D-branes to this exact CFT background, we have shown that the mass-deformed supersymmetric Yang-Mills action is reproduced. This follows naturally from the universal structure of the orbifold in which the would-be projected out states in the pure $\N=2$ theory are now projected in the spectrum due to the free action of the orbifold.

Furthermore, we have extended our analysis to the instanton sector by including the appropriate D-branes which engineer the gauge theory instantons. We have shown how the mass-deformed ADHM action naturally follows from our construction and we have written it in a `covariant' form reminiscent of the $\Omega$-deformation. Indeed, it carries the same structure and, in fact, one can derive it from the latter by carefully exchanging space-time fields with internal ones. This is not surprising since, geometrically, the mass and $\Omega$-deformations have the same structure.

Our analysis suggests that this is \emph{all} that one can do while preserving  $\N=2$ supersymmetry and Poincaré invariance\footnote{We do not consider non-commutative deformations which would correspond to turning on some B-field.} of the gauge theory. Not only are these deformations universal, but it has become clear, from the string theory point of view, that there is no room for more inequivalent deformations. We are thus led to conjecture that all gauge theory deformations preserving $\N=2$ and that are described in string theory by an exact CFT are of the type $m$ and $\Omega$, and can be thought of as freely acting orbifolds of $\N=4$ string backgrounds.

Hence, it would be interesting to see whether one can engineer the $\Omega$-deformation and derive the corresponding partition function in string theory directly through a freely-acting orbifold, \emph{i.e.} without resorting to the graviphoton background. In addition, from our experience with the $\Omega$-deformation, one should be able to see precisely how the $\bar m$ dependence is a $Q$-exact deformation of the effective action, therefore justifying the fact that all physical quantities are holomorphic in $m$, at least in the field theory limit, and that one can effectively work to the leading, holomorphic order in the adjoint mass. Besides, it would be illuminating to analyse both $\Omega$ and $m$ at the same time from the orbifold perspective and clarify some subtleties of the $\N=4$ limit in the presence of the $\Omega$-deformation\footnote{We thank F. Morales for pointing this to us.}.

Beyond the gauge theory level, one wonders whether holomorphy is preserved by $\alpha'$ corrections. Perturbatively, this can be clearly derived from the exact perturbative expressions calculated in \cite{Florakis:2015ied}, and it would be interesting to analyse the same question non-perturbatively using topological string methods \cite{Bershadsky:1993cx,Antoniadis:1993ze,Antoniadis:2015spa}.


\section*{Acknowledgements}

We thank M.~Frau, P.~Larocca, C.~Maccaferri and I.~Pesando for discussions.
M.M. wishes to thank the ICTP, Trieste and A.Z.A. would like to thank the CERN Theory Department and the Physics Department of the University of Torino for their warm hospitality during the accomplishment of this work.
The work of M.M. was partially supported by the Compagnia di San Paolo 
contract ``MAST: Modern Applications of String Theory'' TO-Call3-2012-0088.

\appendix

\section{Notations and conventions}\label{appendix:spinors}
\subsection{Spinors}

In this appendix, we present some of our notations and conventions. $SO(4)$ spinor indices are raised and lowered using the epsilon-tensors
\begin{align}
 &\epsilon^{12}=\epsilon_{12} = 1\,,\,\epsilon^{\dot{1}\dot{2}}=\epsilon_{\dot{1}\dot{2}} = -1\,,
\end{align}
such that
\begin{align}
&\psi^\alpha = +\epsilon^{ab}\psi_b \,,&&\psi_a=-\epsilon_{ab}\psi^b\,,&&\psi^{\dot a} = -\epsilon^{\dot a\dot b}\psi_{\dot b} \,,&&\psi_{\dot a}=+\epsilon_{\dot a\dot b}\psi^{\dot b}\,.
\end{align}
In addition, the $SO(4)$ $\sigma$-matrices $(\sigma^k)_{a\dot a}$ and $(\bar\sigma^k)^{\dot aa}$ are defined as
\begin{align}
	\sigma^\mu = (1\!\!1,-i\boldsymbol{\sigma})\,, && \bar\sigma^k = (1\!\!1,+i\boldsymbol{\sigma})\,,
\end{align}
and are related to one-another by transposition.
On the other had, we denote $S_{\alpha}$ (resp. $S_{\dot\alpha}$ the self-dual (resp. anti-self-dual) spin fields of the space-time $SO(4)$. The spin fields for the internal manifold are, instead, $S_A$, $S^A$, $S_{\hat A}$, $S^{\hat A}$. Notice that covariant and contravariant indices $(A,\hat{A})$ of $SO(2)_{\pm}$ reflect charges $\pm 1/2$ with respect to $SO(2)$ according to the decomposition $SO(6)\rightarrow SO(2)\times SO(4)$ so that care must be taken regarding their position. Our conventions for the internal spin fields can be found in the table below.
\begin{align}\renewcommand{\arraystretch}{1.4}
	\begin{array}{c |c| c}
					\textrm{Spin field}  & SO(2) & SO(4) \\ \hline \hline
					S_A & - & (--),(++)\\ \hline
					S^A & + & (++),(--) \\ \hline
					S_{\hat{A}} & + & (-+),(+-) \\ \hline 
					S^{\hat{A}} & - & (+-),(-+) \\ \hline
	\end{array}\label{TabConventions}
\end{align}

\subsection{Operator product expansions}\label{OPEs}

The operator product expansion algebra for the ten-dimensional fields can be decomposed according to the compactified theory. Indeed, the space-time current algebra is
\begin{alignat}{5}
&S^{\dot\alpha}(z)S_{\beta}(w)\,&\sim&\,\frac{1}{\sqrt{2}}\,{(\bar\sigma^\mu)^{\dot\alpha}}_{\beta}\psi_\mu(w)\,,\,&S_{\alpha}(z)S^{\dot\beta}(w)\,&\sim\,\frac{1}{\sqrt{2}}\,{(\sigma^\mu)_{\alpha}}^{\dot\beta}\psi_\mu(w)\,,\\
&S^{\dot\alpha}(z)S^{\dot\beta}(w)\,&\sim&\,-\frac{\epsilon^{\dot\alpha\dot\beta}}{(z-w)^{1/2}}\,,\,&S_{\alpha}(z)S_{\beta}(w)\,&\sim\,\frac{\epsilon_{\alpha\beta}}{(z-w)^{1/2}}\,,\\
&\psi^{\mu}(z)S^{\dot\alpha}(w)\,&\sim&\,\frac{1}{\sqrt{2}}\frac{(\bar\sigma^\mu)^{\dot\alpha\beta}S_{\beta}(w)}{(z-w)^{1/2}}\,,\,&\psi^{\mu}(z)S_{\alpha}(w)\,&\sim\,\frac{1}{\sqrt{2}}\frac{(\sigma^\mu)_{\alpha\dot\beta}S^{\dot\beta}(w)}{(z-w)^{1/2}}\,,\\
&J^{\mu\nu}(z)S^{\dot\alpha}(w)\,&\sim&\,-\frac{1}{2}\frac{{(\bar\sigma^{\mu\nu})^{\dot\alpha}}_{\dot\beta}S^{\dot\beta}(w)}{z-w}\,,\,\,\,&J^{\mu\nu}(z)S_{\alpha}(w)\,&\sim\,-\frac{1}{2}\frac{{(\sigma^{\mu\nu})_{\alpha}}^{\beta}S_{\beta}(w)}{z-w}\,,
\end{alignat}
whereas the internal one is given by
\begin{alignat}{5}
&S^{A}(z)S_{B}(w)\,&\sim&\,\frac{i{\delta^{A}}_{B}}{(z-w)^{3/4}}\,,\,&S_{A}(z)S^{B}(w)\,&\sim\,\frac{i{\delta_{A}}^{B}}{(z-w)^{3/4}}\\
&S^{A}(z)S^{B}(w)\,&\sim&\,\frac{i}{\sqrt{2}}\frac{(\Sigma^I)^{AB}\psi_{I}}{(z-w)^{1/4}}\,,\,&S_{A}(z)S_{B}(w)\,&\sim\,-\frac{i}{\sqrt{2}}\frac{(\Sigma^I)_{AB}\psi_{I}}{(z-w)^{1/4}}\,,\\
&\psi^{I}(z)S_{A}(w)\,&\sim&\,\frac{1}{\sqrt{2}}\frac{(\bar\Sigma^I)_{AB}S^{B}(w)}{(z-w)^{1/2}}\,,\,&\psi^{I}(z)S^{A}(w)\,&\sim\,-\frac{1}{\sqrt{2}}\frac{(\Sigma^I)^{AB}S_{B}(w)}{(z-w)^{1/2}}\,,\\
&J^{IJ}(z)S^{A}(w)\,&\sim&\,\frac{1}{2}\frac{{(\bar\Sigma^{IJ})^{A}}_{B}S^{B}(w)}{z-w}\,,\,\,\,&J^{IJ}(z)S_{A}(w)\,&\sim\,\frac{1}{2}\frac{{(\Sigma^{IJ})_{A}}^{B}S_{B}(w)}{z-w}\,.
\end{alignat}

Using the algebras above, one easily derives all the necessary correlation functions used troughout the manuscript. Finally, a useful property is the two-point function for fields with Dirichlet boundary conditions. Indeed, using
\begin{equation}
 \left\<Z_{\textrm{DD}}(z_1)\bar Z_{\textrm{DD}}(z_2)\right\>=\log|z_{12}|^2-\log|z_{1\bar2}|^2\,,
\end{equation}
we find
\begin{align}
 \left\<\Pa Z_{\textrm{DD}}(z_1)\Pa \bar Z_{\textrm{DD}}(z_2)\right\>=&\frac{1}{z_{12}^2}\,,\\
 \left\<\Pa Z_{\textrm{DD}}(z_1)\Bp \bar Z_{\textrm{DD}}(z_2)\right\>=&-\frac{1}{z_{1\bar2}^2}\,,\\
 \left\<\Bp Z_{\textrm{DD}}(z_1)\Bp \bar Z_{\textrm{DD}}(z_2)\right\>=&\frac{1}{z_{\bar1\bar2}^2}\,.
\end{align}
For the N-N directions, the same result holds with positive sign for all the two-point functions. For the fermions in the NS sector, it is the opposite:
\begin{align}
 \left\<\psi_{NN}(z_1)\psi_{NN}(z_2)\right\>=&\frac{1}{z_{12}}\,,&\,\left\<\psi_{NN}(z_1)\psi_{NN}(\bar z_2)\right\>=&-\frac{1}{z_{1\bar2}}\,,\\
 \left\<\psi_{DD}(z_1)\psi_{DD}(z_2)\right\>=&\frac{1}{z_{12}}\,,&\,\left\<\psi_{DD}(z_1)\psi_{DD}(\bar z_2)\right\>=&\frac{1}{z_{1\bar2}}\,.
\end{align}

\section{Undoing the orbifold}\label{appendix:vertex}

\subsection{Vertex operator}

Instead of working directly with the $\sigma$-model deformation \eqref{eq:deltaS}, and since our goal is to reproduce the mass-deformed effective actions using a particular vertex operator, one may guess the latter from the physical picture of the deformation and using the form of the non-trivial metric \eqref{eq:metric} of the corresponding Melvin background. For this, we consider the `internal' part of the graviton
\begin{equation}\label{Vgrav}
V_{g}=h_{IJ}\left(\partial Z^I+i(p\cdot \psi)\psi^I\right)\left(\bar\partial Z^J+i(p\cdot\tilde\psi)\tilde\psi^J\right)e^{i\,p\cdot Z}\,,
\end{equation}
and identify the mass parameter as a constant background for this operator in some of the internal directions. More precisely, $h_{3k}$, $h_{3\bar k}$, $h_{\bar3k}$ and $h_{\bar3\bar k}$ are identified with the mass to linear order in the momenta along $\mb C^2$ while $h_{33}$, $h_{3\bar3}$ and $h_{\bar3\bar3}$ give the quadratic mass deformation at quadratic momentum order. For instance, along the $3,k$ direction, the vertex operator is given a background value of $(-)^k m/4\pi$.

In order to test this prescription, we calculate all tree-level (disc) amplitudes between the gauge theory (resp. ADHM) fields and the mass operator to leading order in $\alpha'$ \cite{Billo:2006jm,Ito:2009ac,Ito:2010vx}.  In this way, we obtain a mass-deformed version of the Yang-Mills (resp. ADHM) action and verify that it matches the one obtained from the freely-acting orbifold construction.

\subsection{Perturbative sector}

\begin{figure}[h!t]
\begin{center}
-\includegraphics[width=9cm]{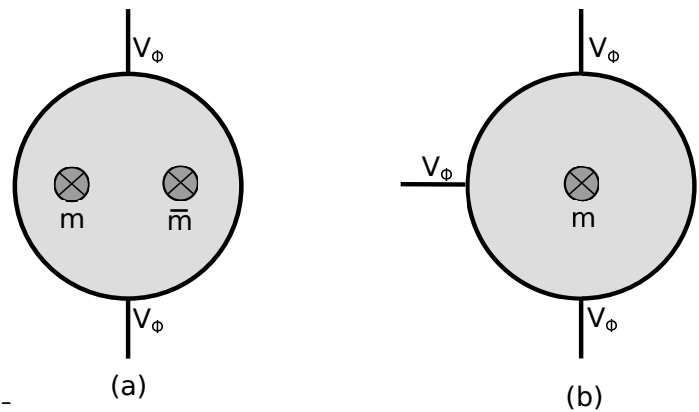}
\end{center}
\parbox{\textwidth}{\caption{Scalar amplitudes with the closed string state of the mass deformation. (a) contributes to the mass of $\phi$ while (b) gives the trilinear coupling.}\label{DiagYM}}
\end{figure}

We first analyse the gauge sector. The only potentially non-vanishing diagrams in the $\alpha'\rightarrow0$ limit involve two scalars and one or two mass vertices, or three scalars and one mass vertex. These are schematically depicted in Fig. \ref{DiagYM} where the boundary of the disc lies completely on a D5-brane.

Instead of calculating directly the coupling of the full `graviton' operator to the gauge theory fields, it is illuminating to expand it in the mass and consider separately the contributions of the linear and quadratic parts. In this way, as shown below, we see that the two parts conspire to give a mass to $\phi_k$ while they cancel against each other to keep $\chi$ massless. Hence, the importance of the quadratic deformation is elucidated.

Let us first focus on the scalar $\chi$ and consider the term $m^2\bar\chi^2$ which receives contributions from two different amplitudes, namely $\left<V_{33}(z,\bar z)V_{\bar\chi}(x_1)V_{\bar\chi}(x_2)\right>$ and $\left<V_{3k}(z_1,\bar z_1)V_{3\bar k}(z_2,\bar z_2)V_{\bar\chi}(x_1)V_{\bar\chi}(x_2)\right>$. We first consider the former amplitude in which we fix the positions $z,\bar z$ and $x_1$ below to $i,-i,-\infty$ and use the scalar vertices in the $-1$ picture. Hence, $x_2$ is integrated on the full real line and the calculation goes as follows:

\begin{align}
 \left\<\left\<V_{33}(z,\bar z)V_{\bar\chi}(x_1)V_{\bar\chi}(x_2)\right\>\right\>=&\frac{m^2\bar\chi^2}{16\pi}\sum_{k=1,2}\int_{\mb R}\textrm{d}x_2\,(z-\bar z)|z-x_1|^2\notag\\
 &\times\left\<\left(\psi^k\psi^3(z)\bar\psi^k\psi^3(\bar z)+\bar\psi^k\psi^3(z)\psi^k\psi^3(\bar z)\right)\bar\psi^3 e^{-\varphi}(x_1)\bar\psi^3 e^{-\varphi}(x_2)\right\>\notag\\
 =&\frac{m^2\bar\chi^2}{4\pi}\int_{\mb R}\textrm{d}x_2\frac{|z-x_1|^2}{x_{12}}\left\<\psi^3(z)\psi^3(\bar z)\bar\psi^3(x_1)\bar\psi^3(x_2)\right\>\,.\label{XXchichi}
\end{align}
The fermionic correlator is straightforward and, integrating over the position $x_2$, we find
\begin{align}
 \left\<\left\<V_{33}(z,\bar z)V_{\bar\chi}(x_1)V_{\bar\chi}(x_2)\right\>\right\>=&\frac{i}{2}m^2\bar\chi^2\,.
\end{align}

We now turn to the second contribution in which we fix $z_1,\bar z_1,x_1$ to $i,-i,\infty$. On easily finds
\begin{align}
\left\<\left\<V_{3k}(z_1,\bar z_1)V_{3\bar k}(z_2,\bar z_2)V_{\bar\chi}(x_1)V_{\bar\chi}(x_2)\right\>\right\>=&-\frac{m^2\bar\chi^2}{32\pi^2}\int_{\mb C\times\mb R}\textrm{d}^2z_2\textrm{d}x_2 z_{1\bar1}\frac{|z_1-x_1|^2}{x_{12}}\notag\\
&\times\sum_{k=1,2}	\left\<(\bar\psi^k\psi^3(z_1)\Bp Z^k(\bar z_1)\psi^k\psi^3(z_2)\Bp\bar Z^k(\bar z_2)\bar\psi^3(x_1)\bar\psi^3(x_2)\right\>\notag\\
&+(1,2)\leftrightarrow(\bar1,\bar2)\notag\\
=&\frac{m^2\bar\chi^2}{16\pi^2}\int_{\mb C\times\mb R}\textrm{d}^2z_2\textrm{d}x_2 \frac{z_{1\bar1}}{z_{\bar1\bar2}^2(z_1-x_2)(z_2-x_2)}+(1,2)\leftrightarrow(\bar1,\bar2)\,.\label{XkXkchichi}
\end{align}
Integrating over $x_2$, and using the elementary integral
\begin{equation}\label{ElemInt}
 \int_{\mb H_-}\textrm{d}^2z\frac{y-\bar y}{(z-y)(\bar z-\bar y)^2}=\pi\,,
\end{equation}
we obtain
\begin{align}
\left\<\left\<V_{3k}(z_1,\bar z_1)V_{3\bar k}(z_2,\bar z_2)V_{\bar\chi}(x_1)V_{\bar\chi}(x_2)\right\>\right\>=&-\frac{i}{2}m^2\bar\chi^2\,.
\end{align}

Therefore, the coupling $m^2\bar\chi^2$ does not appear in the effective action. Let us now turn to the more interesting term $|m|^2|\chi|^2$. The calculation goes exactly along the same lines as the previous one and one finds that the total mass term for $\chi$ is zero.

We now calculate the masses for the other two scalars. Notice that there can be no terms such as $m^2\bar\phi_k^2$ or its complex conjugate because of the particular structure of the mass vertex operators. In addition, as we shall see, the same cancellation occurring for the $\chi$ terms again shows up. However, there are additional contributions from the linear deformation leading to a non-zero term. Let us see this more precisely by first considering the quadratic deformation. Since the calculation goes as the one in \eqref{XXchichi}, we simply give the result:

\begin{align}
 \left\<\left\<V_{3\bar 3}(z,\bar z)V_{\phi_l}(x_1)V_{\bar\phi_m}(x_2)\right\>\right\>=&\frac{i}{4}|m|^2\delta^{lm}\phi_l\bar\phi_m\,.
\end{align}

As for the term arising from the linear deformation, we notice that, contrarily to the case of $\chi$, all the terms containing fermions in linear part of \eqref{Vgrav} contribute. We call the terms in the direction $(k,3)$ non-diagonal and the ones in $(k,k)$ the diagonal ones. Let us first focus on the non-diagonal terms, where the mass vertices must carry the same index $k$ to lead to a non-trivial result. Referring to \eqref{XkXkchichi} for the technical details, we find
\begin{align}
 \left\<\left\<V_{3k}(z_1,\bar z_1)V_{\bar 3\bar k}(z_2,\bar z_2)V_{\phi_l}(x_1)V_{\bar\phi_m}(x_2)\right\>\right\>_{\textrm{nd}}=&-\frac{i}{8}|m|^2\delta^{lm}\phi_l\bar\phi_m\,.
\end{align}
The other non-diagonal contribution stems from the $V_{X\bar k},V_{\bar Xk}$ vertices and is exactly equal. Hence, as announced above, the non-diagonal terms cancel the contribution from the quadratic deformation, and we can focus on the diagonal terms only:

\begin{align}
 \left\<\left\<V_{3k}(z_1,\bar z_1)V_{\bar 3\bar k}(z_2,\bar z_2)V_{\phi_l}(x_1)V_{\bar\phi_m}(x_2)\right\>\right\>_{\textrm{d}}&=\frac{|m|^2\phi_l\bar\phi_m}{32\pi^2}\int_{\mb C\times\mb R}\textrm{d}^2z_2\textrm{d}x_2 z_{1\bar1}|z_1-x_1|^2\notag\\
\times&\sum_{k=1,2}\left\<\left(\bar\psi^k\psi^k(z_1)\Bp X(\bar z_1)+\Pa X(z_1)\bar\psi^k\psi^k(z_1)\right)\right.\notag\\
\times&\left.\left(\psi^k\bar\psi^k(z_2)\Bp\bar X(\bar z_2)+\Pa\bar X(z_2)\psi^k\bar\psi^k(z_2)\right)\psi^l e^{-\varphi}(x_1)\bar\psi^m e^{-\varphi}(x_2)\right\>\notag\\
=&-\frac{|m|^2|\phi_l\bar\phi_m}{8\pi^2}\int_{\mb C\times\mb R}\textrm{d}^2z_2\textrm{d}x_2 z_{1\bar1}\frac{|z_1-x_1|^2}{x_{12}}\notag\\
&\times\left\<(\bar\psi^k\psi^k(z_1)\Bp X(\bar z_1)\psi^k\bar\psi^k(z_2)\Bp\bar X(\bar z_2)\psi^l(x_1)\bar\psi^m(x_2)\right\>\,.
\end{align}
Therefore, we find
\begin{align}
 \left\<\left\<V_{3k}(z_1,\bar z_1)V_{\bar 3\bar k}(z_2,\bar z_2)V_{\phi_l}(x_1)V_{\bar\phi_m}(x_2)\right\>\right\>_{\textrm{d}}=&-\frac{i}{4}|m|^2\delta^{lm}\phi_l\bar\phi_m\,,
\end{align}
and one obtains a mass for $\phi_k$. We now turn to the trilinear couplings. There are four possible terms:
\begin{align}
 \left\<V_{3k}V_{\bar\chi}V_{\phi_l}V_{\bar\phi_m}\right\>\,,&&\,\left\<V_{3\bar k}V_{\bar\chi}V_{\phi_l}V_{\bar\phi_m}\right\>\,,&&\,\left\<V_{\bar 3k}V_{\chi}V_{\phi_l}V_{\bar\phi_m}\right\>\,,&&\,\left\<V_{\bar 3\bar k}V_{\chi}V_{\phi_l}V_{\bar\phi_m}\right\>\,.
\end{align}
We choose to fix the positions of the scalars (\emph{e.g.} at $0,1,\infty$) and integrate over the position of the closed string vertex. The first term yields
\begin{align}
 \left\<\left\<V_{3k}(z,\bar z)V_{\bar\chi}(x_1)V_{\phi_l}(x_2)V_{\bar\phi_m}(x_3)\right\>\right\>=&\frac{i(-)^km\bar\phi_3\phi_l\bar\phi_m}{4\pi}\int_{\mb C}\textrm{d}^2z\left\<\left(\Pa X(z)\bar\psi^k\psi^k(\bar z)+\bar\psi^k\psi^k(z)\Bp X(\bar z)\right)\right.\notag\\
 \times&x_{12}x_{13}x_{23}\left.\left(\Pa\bar X-i(p\cdot\psi)\bar\chi\right)(x_1)\psi^le^{-\varphi}(x_2)\bar\psi^me^{-\varphi}(x_3)\right\>\notag\\
 =&\frac{i(-)^{k+1}m\bar\phi_3\phi_l\bar\phi_m}{4\pi}\int_{\mb C}\textrm{d}^2z\frac{x_{12}x_{13}}{(z-x_1)^2}\left\<\bar\psi^k\psi^k(\bar z)\psi^l(x_2)\psi^m(x_3)\right\>\notag\\
 =&\frac{i(-)^{k}m\bar\phi_3\phi_l\bar\phi_m\delta^{kj}\delta^{kl}}{4\pi}\int_{\mb C}\textrm{d}^2z\frac{x_{12}}{(z-x_1)^2(\bar z-x_2)}\notag\\
 =&i(-)^{k+1}m\delta^{kj}\delta^{kl}\bar\phi_3\phi_j\bar\phi_l\,.
\end{align}
The second correlator is exactly equal and the remaining two are similar (they are essentially obtained by complex conjugation). Including the other ordering of $j,l$ and summing over $k$ leads to the correct trilinear coupling.

Let us now focus on the disc diagrams involving the fermionic fields. For simplicity, we start with the full vertex operator \eqref{Vgrav}. Without loss of generality, we focus on the chiral fermions for which the only relevant amplitude is
\begin{align}
 \left\<V_h(z,\bar z)V_\Lambda(x)V_\Lambda(y)\right\>=-ip_m\epsilon_{\alpha\beta}h_{ij}(p)\Lambda^{\alpha A}\Lambda^{\beta B}\frac{(x-y)^{\frac{1}{4}}(z-x)(z-y)}{(\bar z-x)^{\frac{1}{2}}(\bar z-y)^{\frac{1}{2}}}\left\<\psi^{m}\psi^{i}(z)\tilde\psi^j(\bar z)S_A(x)S_B(y)\right\>\,,
\end{align}
where the left-right symmetrisation is implicitly understood. In order to evaluate the remaining correlator, it is sufficient to consider the OPEs (see Appendix \ref{OPEs}) of the fermion bilinear at $z$ and the spin fields at $x$ and $y$ and then evaluate the integral over the unfixed position. Indeed, the CFT correlator leads to
\begin{align}
 \frac{(x-y)^{\frac{1}{4}}(z-x)(z-y)}{(\bar z-x)^{\frac{1}{2}}(\bar z-y)^{\frac{1}{2}}}\left\<\psi^{m}\psi^{i}(z)\tilde\psi^j(\bar z)S_A(x)S_B(y)\right\>\sim -\frac{i}{2\sqrt{2}}{(\Sigma^{[mi})_B}^A(\bar\Sigma^{\tilde j]})_{AC}\frac{x-y}{(\bar z-x)(\bar z-y)}\,.
\end{align}
All in all, we obtain
\begin{align}
 \left\<\left\<V_h(z,\bar z)V_\Lambda(x)V_\Lambda(y)\right\>\right\>&=-i\sqrt{2}p_m(\Sigma^{mi\tilde j})_{AB} h_{ij}(p)\Lambda^{\alpha A}{\Lambda_{\alpha}}^B\notag\\
 &=2\sqrt{2}i\,m\left(\Sigma^{1\bar13}-\Sigma^{2\bar23}\right)_{AB}\Lambda^{\alpha A}{\Lambda_{\alpha}}^B\,.
\end{align}
Putting all terms together, the effective action \eqref{SYMaction} is recovered.

\subsection{Non-perturbative sector}

\begin{figure}[h!t]
\begin{center}
\includegraphics[width=9cm]{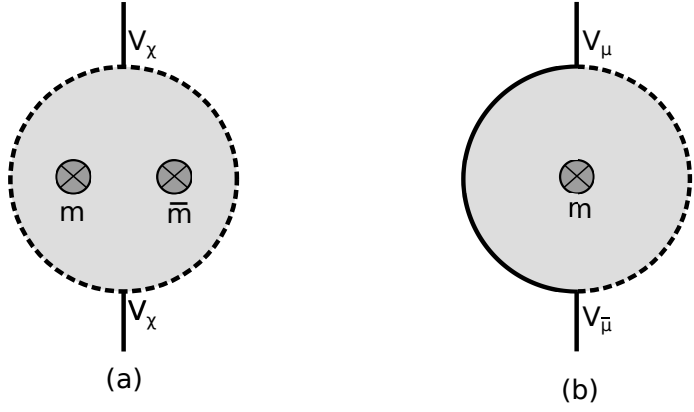}
\end{center}
\parbox{\textwidth}{\caption{Disc amplitudes with the closed string state of the mass deformation in the instanton sector. (a) has its full boundary on the D-instantons and contributes to the mass of the bosonic ADHM moduli, whereas (b) has a mixed boundary on the D5-branes and the D1-branes and contributes to the mass of the fermionic mixed moduli.}\label{DiagInst}}
\end{figure}

The relevant disc diagrams are depicted in Fig. \ref{DiagInst}.
Here, things are slightly more complicated because of the existence of the mixed states. However, technically, the discussion goes along the same lines. Notice that, at leading order in $\alpha'$, we cannot scatter mixed and unmixed states since at least two mixed states are necessary to have a non-trivial amplitude.

Let us first focus on the amplitudes with unmixed states. Similarly to the Yang-Mills sector, we complexify and split the scalars into $\chi_I$ into $\phi$ and $\chi_k$ with $k=1,2$:
\begin{align}
 V_{\phi}(z)&=\frac{\phi}{\sqrt{2}}\psi^3(z)e^{-\varphi(z)}\sim\frac{\phi}{\sqrt{2}}\left(\Pa X-i(p\cdot\psi)\psi^3\right)\,,\\
 V_{k}(z)&=\frac{\chi_k}{\sqrt{2}}\psi^k(z)e^{-\varphi(z)}\sim\frac{\chi_k}{\sqrt{2}}\left(\Pa Z^k-i(p\cdot\psi)\psi^k\right)\,.
\end{align}
For the bosonic moduli, it is clear that one obtains the same term as for the Yang-Mills case, simply by replacing the vector multiplet scalar with the ADHM one. Indeed, the only technical difference is the absence of momenta which, however, do not play a particular role in the calculation. Consequently, the modulus $\phi$ stays massless since
\begin{align}
 \left\<\left\<V_{X\bar X}(z,\bar z)V_{\chi}(x_1)V_{\bar\chi}(x_2)\right\>\right\>+\sum_{k,l\in\left\lbrace1,2,\bar1,\bar2\right\rbrace}\left\<\left\<V_{Xk}(z_1,\bar z_1)V_{\bar Xl}(z_2,\bar z_2)V_{\chi}(x_1)V_{\bar\chi}(x_2)\right\>\right\>=0\,.
\end{align}
The moduli $\chi_k$ acquire, instead, a mass term
\begin{align}
 \left\<\left\<V_{X\bar X}(z,\bar z)V_{k}(x_1)V_{\bar k}(x_2)\right\>\right\>+\sum_{l,m\in\left\lbrace1,2,\bar1,\bar2\right\rbrace}\left\<\left\<V_{Xl}(z_1,\bar z_1)V_{\bar Xm}(z_2,\bar z_2)V_{k}(x_1)V_{\bar k}(x_2)\right\>\right\>\sim |m|^2|\chi_k|^2\,.
\end{align}
Finally, the mass deformation generates a trilinear coupling from the following typical disc diagram:
\begin{align}
 \left\<\left\<V_{Xk}(z,\bar z)V_{\bar\chi}(x_1)V_{l}(x_2)V_{\bar m}(x_3)\right\>\right\>=&i(-)^{k+1}m\,\delta^{kj}\delta^{kl}\bar\chi_3\chi_j\bar\chi_l\,.
\end{align}
The same goes for the unmixed fermionic moduli whose vertex operators are exactly of the form of the $\mN=4$ gaugini. Indeed, for the chiral moduli, we find
\begin{align}
 \left\<\left\<V_h(z,\bar z)V_M(x)V_M(y)\right\>\right\>&=2\sqrt{2}i\,m\left(\Sigma^{1\bar13}-\Sigma^{2\bar23}\right)_{AB}M^{\alpha A}{M_{\alpha}}^B\,,
\end{align}
and similarly for the anti-chiral ones.
We now focus on the mixed moduli. It is clear that one needs to insert two mixed moduli so we only have at leading order two possible amplitudes. The ones with bosonic mixed moduli, \emph{e.g.}
\begin{equation}
 \left\<V_h\,V_\omega\,V_{\bar\omega}\right\>\,,
\end{equation}
vanish because the bosonic mixed moduli carry the same space-time chirality. The only possible non-vanishing term is, hence, the one involving the fermionic mixed moduli, \emph{i.e.}
\begin{equation}
 \left\<V_h\,V_\mu\,V_{\bar\mu}\right\>\,.
\end{equation}
Notice that the only difference with the perturbative case is the presence of the twist fields. However, they can only contract between themselves, and the remaining correlator is easily shown to be exactly the same as the one obtained in the fermionic sector in the previous section. Consequently, the result is
\begin{align}
 \left\<V_h\,V_\mu\,V_{\bar\mu}\right\>=2\sqrt{2}i\,m\left(\Sigma^{1\bar13}-\Sigma^{2\bar23}\right)_{AB}\mu^{A}\bar\mu^B\,.
\end{align}

\bibliographystyle{utphys}
\bibliography{references}
\end{document}